\documentclass{llncs}
\usepackage{latexsym,epsfig,graphicx,amsmath}
\title{Short Report on: Possible directions to Auctions with Cryptographic pre-play}
\date{}
\author{Amjed Shareef\\
Department of Computer Science and Engineering,\\
Indian Institute of Technology Madras, Chennai - 600036, India.\\
Email:  \tt amjedshareef@gmail.com \\ }
\institute{}
\begin{document}
\maketitle
\begin{abstract}
In auction theory, cryptography has been used to achieve anonymity of the participants, security and privacy of the bids, secure computation and to simulate mediator (auctioneer). Auction theory focuses on revenue and Cryptography focuses on security and privacy.   Involving Cryptography at base level, to enhance revenue gives entirely new perspective and insight to Auction theory, thereby achieving the core goals of auction theory.
In this report, we try to investigate an interesting field of study in Auction Theory using Cryptographic primitives.\\
\end{abstract}
\section{Introduction}

\subsection{Cryptography Primitives}
Yao's Millionaires protocol\cite{yao} is the fundamental protocol of modern Cryptography, used by two parties to compare their secret value without third party intervention. For example, if two millionaires want to know which of them is richer without revealing their actual wealth to each other, or any other third person, Yao's protocol can be used. This problem is analogous to a more general problem where there are two numbers $a$ and $b$ and the goal is to solve the inequality $a \geq b$ without revealing the actual values of $a$ and $b$.

\subsection{Auctions theory Basics}
\noindent  \textbf{A first-price sealed-bid auction:}\\
A first-price sealed-bid auction is a form of auction where bidders submit one bid in a concealed fashion. The submitted bids are then compared and the person with the highest bid wins the award, and pays the amount of his bid to the seller. \\

\noindent \textbf{Asymmetric bidders}\\
If the  the probability distribution from which the bidders obtain their values is identical across bidders, then they are symmetric bidders, 
otherwise asymmetric bidders.\\

\noindent  \textbf{Related work in Auction theory}\\
Landsberger et.al\cite{sim} in the paper "First-Price Auctions when the Ranking of Valuations is Common Knowledge" studied a one unit two bidders game where the valuations ranking is known. At the beginning of the game, the valuations are drawn from a distribution that is common knowledge and the seller announces who is the bidder with the highest valuation. This set-up induces a particular asymmetric auction model  and proved existence and uniqueness of equilibrium in pure strategies in the two-bidder case.

Philippe\cite{who} considered  a sequential auction where two identical goods are sold sequentially via first price auctions to players who are interested in both objects.  In sealed bid auctions, no information is a priori revealed by the mechanism, but the seller can in principle reveal whatever he wants. They have considered the case where,  only the information revealed between the two stages of the game is the name of the winner in the first auction. 
They have proved that, the seller must not reveal the name of the winner between the two auctions and instead sell both goods using a simultaneous auction,
 to get more revenue. 

\section{Auctions with Cryptographic pre-play}
Now, let us consider a situation where, there are two parties who are going to bid in first price sealed bid auction.
We assume that parties have interest to compare their bids.  If these two parties, prior to actual auction (say trail auction), compare their bid using Yao' protocol, to know who's bid is higher, then few interesting question arise, \\
What will they bid in actual auction?\\
Will it increase or decrease overall revenue of actual auction?\\

\noindent  The solution seems to be trivial. The looser in trail auction will increase his bid in actual action, and winner will decrease his bid in actual auction. It is important to note that the above information is common knowledge to both the parties. Considering this fact, the looser may not increase his bid assuming that the winner is going to decrease(or may not increase)  his bid and vice versa. To solve this we need to adopt strong auction theory model.
An asymmetry arises here because the players who participate in the actual auction already participated in the trail auction. The information given at the end of the trail auction is used by the bidders to revise their beliefs and the actual auction must be studied under the hypothesis that bidders are asymmetric. Sometimes it must be assumed that the distribution function is no longer common knowledge.\\
Basically, we have  to answer the following questions.\\
What will the player bid in trial auction?\\
What is the equilibrium of trail auction?  \\

Players may fake their bid in the trail auction, if it is going to increase their pay off (revenue) in actual auction. Participating in trial auction with the honest bid gives some information about other party's bid, but at the same time leaks the information about their own bid. 

\section{Conclusion}
Currently we are working on this problem.  Even though, we could not conclude anything concretely yet, it is interesting to study the effects of using cryptography protocols in Auction Theory. Any change in auction revenue, with Cryptographic pre-play has significant impact on Auction theory. 

\bibliography{auction}
\end{document}